\documentclass[amsmath,amsfonts,twocolumn,showpacs,superscriptaddress,prb]{revtex4}
\usepackage{graphicx}
\usepackage{epsfig}
\usepackage{bbm}
\usepackage{bm}
\usepackage{dcolumn}
\usepackage{amsmath}
\usepackage{amssymb}
\usepackage{color}
\usepackage[varg]{txfonts}

\newcommand{\beg}{\begin{equation}}
\newcommand{\en}{\end{equation}}

\newcommand{\bq}{\mathbf q}

\newcommand{\br}{\mathbf r}

\newcommand \bel  {\begin{align}}
\newcommand \enl  {\end{align}}

\newcommand{\veps}{\varepsilon}
\newcommand{\eps}{\epsilon}

\def\Xint#1{\mathchoice
   {\XXint\displaystyle\textstyle{#1}}%
   {\XXint\textstyle\scriptstyle{#1}}%
   {\XXint\scriptstyle\scriptscriptstyle{#1}}%
   {\XXint\scriptscriptstyle\scriptscriptstyle{#1}}%
   \!\int}
\def\XXint#1#2#3{{\setbox0=\hbox{$#1{#2#3}{\int}$}
     \vcenter{\hbox{$#2#3$}}\kern-.5\wd0}}

\def\dashint{\Xint-}

\begin{document}

\title{Schmid-Higgs Mode in the Presence of Pair-Breaking Interactions
}

\author{Maxim Dzero}
\affiliation{Department of Physics, Kent State University, Kent, OH 44242, USA}

\author{Alex Kamenev}
\affiliation{School of Physics and Astronomy, University of Minnesota, Minneapolis, MN 55455, USA} \affiliation{William I. Fine Theoretical Physics Institute, University of Minnesota, Minneapolis, MN 55455, USA}


\date{\today}

\begin{abstract}
Collective modes in superconductors provided the first realization of the Higgs mechanism. The transverse Goldstone mode acquires a gap (i.e. a mass) when it hybridizes with the electromagnetic gauge field. The longitudinal Schmid-Higgs mode, on the other hand, is always massive. In conventional BCS theory, its gap is exactly $2\Delta$, coinciding with the excitation threshold for quasiparticles.
Being situated right at the edge of the continuum spectrum it gives rise to peculiar dynamics for the Schmid-Higgs mode. For instance, when suddenly excited at $t=0$, it exhibits algebraically decaying oscillations of the form $\sim \sin(2\Delta t)/{t}^{1/2}$. 
In this study, we explore the behavior of Schmid-Higgs oscillations in the presence of pair-breaking mechanisms, such as magnetic impurities or in-plane magnetic fields. These processes suppress the quasiparticle excitation threshold down to $2\veps_g < 2\Delta$, potentially placing the longitudinal mode within the continuum spectrum. Despite this, we show that the algebraically decaying oscillations persist, taking the form  $\sim \sin(2\veps_g t)/t^2$.
The Schmid-Higgs mode becomes truly overdamped and exponentially decaying only in the gapless superconductors with $\veps_g=0$.

\end{abstract}

\maketitle

\section{Introduction}  

Studies of a collective mode associated with the fluctuations in the amplitude of the complex superconducting order parameter, the so-called longitudinal Schmid-Higgs (SH) mode,\cite{ASchmid,VarmaLit1,VarmaLit2} have 
attracted significant attention from both theoretical and experimental condensed matter communities.\cite{Assa2011,Assa2013,Joerg2018,Shimano2020,KotaHighTc1,KotaHighTc2,Cea2014Raman,Measson2014,Behrle2018,Measson2019,Grasset1,Grasset2} Until recently, experimental progress in elucidating the contribution of the SH mode to various response functions has been hampered by the fact that the SH mode does not couple linearly to the electromagnetic field.  Nevertheless, the substantial advances in ultrafast terahertz (THz) spectroscopy have lead not only to the progress in identifying the SH mode, \cite{Shimano2012,Shimano2013,Shimano2014,THz3,Sherman2015-Disorder,Basov2020,KatoKatsumi2024} but also inspired many theoretical studies, which focused on various fundamental aspects of this phenomenon. It was realized, for example, that in a presence of a supercurrent, the amplitude SH mode can be resonantly excited already in the linear order in electromagnetic field.\cite{Moore2017} The injection of the supercurrent also leads to a nonreciprocal second harmonic generation response, recently observed in NbN superconducting films under THz pulses.\cite{NonReciprocal2020} In addition, several works have addressed the role played by scattering on potential impurities in resonant excitation of the SH mode either in single band or multi-band superconductors.\cite{Cea2014-Disorder1,Cea2016,Silaev2019-Disorder,Yang2020-Disorder2,Seibold2021-Disorder,Haenel2021-Disorder,Yang2022-Disorder,Eremin2023} The physics associated with the Higgs mode in systems which are tuned to a vicinity of the quantum critical point has also been discussed.\cite{Sachdev2012,Dupuis2014}  

Dynamics of the Schmid-Higgs mode in conventional BCS superconductors was extensively addressed in the literature for both spatially homogeneous \cite{VolkovKogan1973,Galaiko1972,Galperin1981,Kulik1981,Spivak2004,Enolski2005} and spatially periodic perturbations. \cite{ChubukovHiggs2023,nosov2024}
The main results of these studies is that at zero momentum ($\bq =0$) energy of the SH mode is exactly $2\Delta_0$ ($\Delta_0$ is the order parameter in a clean superconductor). This puts it right at the threshold of the  continuum spectrum of quasiparticle excitations. Such a coincidence leads to a peculiar dynamics of the SH mode, making it neither freely oscillating, nor exponentially damped. Instead, being excited at, say, $t=0$, the mode exhibits oscillations  with frequency $2\Delta_0$, modulated by an algebraic decay $\sim t^{-1/2}$.\cite{VolkovKogan1973} The maximum of an IR absorption spectrum (e.g., two-photon)  is also expected at exactly $2\Delta_0$ (i.e. $\Delta_0$ per photon), with the spectrum  exhibiting a threshold singularity of the form $(\omega-2\Delta_0)^{-1/2}$.

In this paper we investigate  dynamics of SH mode in superconductors with  pair-breaking mechanisms. The latter are generically associated with a weak breakdown of the Cooper pairing symmetry. In conventional superconductors this is the time-reversal symmetry, which may be broken by, e.g., an in-plane magnetic field (in case of thin film superconductors), or by weak magnetic impurities.\cite{Larkin1965} The effect of such time-reversal symmetry breaking, quantified by a pair-breaking rate, $1/\tau_s$, on the SH mode is of conceptual interest. Indeed, a finite  lifetime of a Cooper pair may be expected to yield an exponential damping of the SH mode even at zero temperature. On the technical level, the pair-breaking processes are known to suppress both the superconducting order parameter  $\Delta$ and the quasiparticle gap  $\veps_g$ in a way that $\veps_g < \Delta <\Delta_0$,\cite{Larkin1965} (in the absence of pair-breaking all three energy scales are equal to each other).  
The ultimate manifestation of this phenomenon is gapless superconductivity, where the order parameter is finite, while the gap in the quasiparticle spectrum is absent. 
Such separation of the two energy scales may lead to the SH mode falling inside the continuum spectrum of quasiparticle excitations, again pointing towards its exponential damping.\cite{Kulik1981} 
However, to the best of our knowledge, no calculation backing up this assertion has been ever presented so far.         

The reason for such an omission despite years of research is in significant technical complexity of the task. One aspect of it stems from the non-perturbative (in the pair-breaking scattering rate, $1/\tau_s$) nature of the result. Indeed, we shall see below that 
$\Delta-\veps_g\sim (1/\tau_s)^{2/3}$ (this aspect was overlooked, e.g., in Ref.~[\onlinecite{DzeroPeriodic2024}]).\cite{Ref41}
Another aspect lies in an intricate analytic structure of the corresponding susceptibility in the complex frequency plane, making it very hard to employ the Matsubara analytical continuation methods, which proved to be useful in the absence of pair-breaking.\cite{Kulik1981,ChubukovHiggs2023,nosov2024} To overcome the latter obstacle, we employ here the real time Keldysh technique.  We work in the disordered limit $\tau\Delta\ll 1$, where $\tau$ is an elastic impurity (non-magnetic) scattering time. This allows us to use the Keldysh sigma-model\cite{Kamenev2009,Kamenev2011} and derive the dynamical Usadel equation,\cite{Usadel1970} generalized to the presence  of pair-breaking processes. We then perform a linear fluctuation analysis of this generalized Usadel equation, along with the proper self-consistency relation, deriving a longitudinal linear susceptibility, $\chi_{\mathrm SH}^R(\omega,\bq)$, of the order parameter. This quantity fully describes dynamics of the SH mode. It must be mentioned that in our subsequent analysis we greatly benefited from the recent paper by Nosov,   Andriyakhina, and Burmistrov,\cite{nosov2024} devoted to the finite $\bq$ properties of the SH mode in conventional superconductors.

The resulting picture is rather intricate. 
In presence of the pair-breaking processes,  the SH peak in the (two-photon) IR absorption spectrum  becomes approximately Lorenzian, centered at the frequency $\omega_{\mathrm{res}}$. This frequency indeed falls within the 
quasiparticle continuum, $\omega_{\mathrm{res}} > \veps_g$. (Notice that the one-photon quasiparticle absorption is absent as long as IR frequency $\omega < 2\veps_g $.) However, 
response to a sudden $t=0$ perturbation is still oscillatory with an {\em algebraic} decay in time, $\sim \sin(2\veps_g t)/t^2$. 
This means that SH mode is still sharply defined at $\omega =2\veps_g$ and is thus robust against relatively weak pair-breaking processes. It becomes exponentially decaying and completely overdamped only in the regime of the gapless superconductivity, where $\veps_g=0$.\cite{Larkin1965,Kamenev2011}


The rest of this paper is organized as follows: in Section II we analyze the generalized Usadel equation and use it to derive the linear dynamical SH susceptibility. In Section III we use the linear susceptibility to analyze the temporal dynamics of SH mode for small momenta. Section IV presents a brief discussion of the results. Some technical details are delegated to two Appendices.

\section{Basic equations}

We consider a conventional superconductor, described by the Bardeen-Cooper-Schrieffer (BCS) Hamiltonian, with broken time reversal symmetry. The latter is achieved by either a parallel magnetic field in case of thin films, or by weak magnetic impurities. 
In order to describe collective excitations in disordered superconductors, we use the Usadel equation\cite{Usadel1970,LarkinVertex,BELZIG1999} for the matrix function $\check{Q}(\br;t_1,t_2)$:\cite{Kamenev2009,Kamenev2011}
\beg\label{Eq1}
\begin{split}
&i\left(\check{\Xi}_3\,\partial_{t_1}\check{Q}+\partial_{t_2}\check{Q}\,\check{\Xi}_3\right)-iD{\vec \partial}_\br\left(\check{Q}\circ{\vec \partial}_\br\check{Q}\right)+\left[\check{\Delta},\check{Q}\right]\\&=-\frac{i}{6\tau_{\textrm s}}\sum\limits_{a=1}^3\left[\left(\hat{\rho}_3\otimes\hat{\sigma}_a\right)\check{Q}\left(\hat{\rho}_3\otimes\hat{\sigma}_a\right)\stackrel{\circ},\check{Q}\right]. 
\end{split}
\en
Here $1/\tau_{\textrm{s}}$ is the pair-breaking  rate, $\hat{\rho}_a$ are the Pauli matrices which act in Nambu space, $\hat{\sigma}_a$ are the Pauli matrices acting in spin space, matrix 
$\check{\Xi}_3=\hat{\gamma}^{\textrm{cl}}\otimes(\hat{\rho}_3\otimes\hat{\sigma}_0)=\hat{\gamma}^{\textrm{cl}}\otimes\hat{\Xi}_3$ is diagonal in Keldysh space, $\hat{\gamma}^{\textrm{cl}}$ is a unit matrix in Keldysh space, $\hat{\rho}_0$ is a unit matrix in Nambu space, $D=v_F^2\tau/d$  is the diffusion coefficient, $\tau$ is the elastic scattering time on potential disorder, $d$ is the system's dimensionality and $\check{A}{\circ}\check{B}$ denotes the usual convolution with respect to time and space variables.

To derive the pair-breaking term on the right hand side of the Usadel equation (\ref{Eq1}), one starts from the magnetic disorder term in the action $\int\! d{\bf r} dt\, \bar\psi({\bf r},t) B_{\mathrm{dis}}^a({\bf r}) \hat{\rho}_3\otimes\hat{\sigma}_a \psi({\bf r},t)$. After averaging over the potential disorder, one expands the 
sigma-model action to the second order in the random magnetic field, $B_{\mathrm{dis}}^a({\bf r})$, and averages over it with the  variance in the form: $\left\langle B_{\mathrm{dis}}^a({\bf r})B_{\mathrm{dis}}^b({\bf r'})\right\rangle=(2\pi\nu \tau_s)^{-1} \,\frac{1}{3}\, \delta^{ab}\,\delta({\bf r}-{\bf r'})$. 
 The Usadel equation is then obtained by a variation of the resulting action over $\delta/\delta \check{Q}({\bf r})$, subject to the constraint $\check{Q}^2=\mathbbm{1}$.\cite{Kamenev2011} The implicit assumption about the weakness of scattering induced by magnetic impurities is reflected in the fact that the averaging over their distribution is performed after the averaging with respect to the distribution of potential impurities.\cite{Austin2000} In other words, account of the scattering due to magnetic disorder is approximate and it is controlled by the small parameter $l/l_s\ll 1$, where $l$ and $l_s$ are the mean free paths for the scattering on potential and magnetic impurities correspondingly.

The superconducting order parameter matrix $\check{\Delta}(\br,t)$ is defined 
according to
$\check{\Delta}(\br,t)=\Delta(\br,t)\left(\hat{\gamma}^{\textrm{cl}}\otimes\hat{\rho}_{+}\otimes\hat{\sigma}_0\right)-\overline{\Delta}(\br,t)\left(\hat{\gamma}^{\textrm{cl}}\otimes\hat{\rho}_{-}\otimes\hat{\sigma}_0\right)$, 
where  $\hat{\rho}_{\pm}=\hat{\rho}_1\pm i\hat{\rho}_2$. Note that $\check{\Delta}$ is diagonal in Keldysh space. The order parameter is to be determined  from the self-consistency relation, obtained through the variation of the action with respect to the quantum component of the $\check\Delta$,  
\beg\label{SelfConsistent}
\Delta(\br,t)=\frac{\pi\lambda}{2}\,\, \textrm{Tr}\left\{\left(\hat{\gamma}^{\textrm{q}}\otimes\hat{\rho}_{-}\otimes\hat{\sigma}_0\right)
\otimes\check{Q}(\br;t,t)\right\}, 
\en
where $\hat{\gamma}^{\textrm{q}}$ is the first Pauli matrix acting in the Keldysh subspace and $\lambda$ is the dimensionless BCS coupling constant.

\subsection{Ground state}

It is convenient to perform the Wigner transformation with respect to the relative time $\tau=t_1-t_2$:
\beg\label{WTeps}
\check{Q}(\br;t_1,t_2)=\int\frac{d\eps}{2\pi}\,\, \check{Q}_{\eps}(\br,t)\, e^{-i\eps(t_1-t_2)}
\en
and $t=(t_1+t_2)/2$. In the ground state, matrix $\check{Q}_\eps$ is static and  spatially homogeneous, provided that there are no externally imposed boundaries. The retarded and advanced components of $\check{Q}_\eps$, denoted as $\check{\Lambda}_\eps^{R(A)}$, can be parametrized as follows:\cite{Kamenev2011,Savich2017}
\beg\label{LepsRA}
\begin{split}
&\hat{\Lambda}_\eps^{R}= \hat{\Xi}_3 \cosh\vartheta_\eps + \hat{\Xi}_2 \sinh\vartheta_\eps\cos\chi_\eps +
i\,\hat{\Xi}_1 \sinh\vartheta_\eps\sin\chi_\eps
\end{split}
\en
and $\hat{\Lambda}_\eps^A=-\left[\hat{\Lambda}_\eps^{R}\right]^\dagger$. In equation (\ref{LepsRA})
matrices are defined according to $\hat{\Xi}_1=\hat{\rho}_1\times\hat{\sigma}_0$ and $\hat{\Xi}_2=i\hat{\rho}_2\times\hat{\sigma}_0$.
In the ground state and in the absence of a current flowing in a superconductor one can set $\chi_\eps=0$. 
It is straightforward to verify that both $\hat{\Lambda}_\eps^{R(A)}$ satisfy the normalization condition 
$\hat{\Lambda}_\eps^R\hat{\Lambda}_\eps^R=\hat{\Lambda}_\eps^A\hat{\Lambda}_\eps^A=\mathbbm{1}$. The {\em complex} Nambu angle, $\vartheta_\eps$, describing the ground state, is then found from the equation 
\beg\label{GSUsadel}
\begin{split}
&\eps\left[\check{\Xi}_3,\hat{\Lambda}_\eps^R\right]+\left[\hat{\Delta},\hat{\Lambda}_\eps^R\right]\\&+\frac{i}{6\tau_{\textrm{s}}}\sum\limits_{a=1}^3\left[(\hat{\rho}_3\times\hat{\sigma}_a)\hat{\Lambda}_\eps^R(\hat{\rho}_3\times\hat{\sigma}_a),\hat{\Lambda}_\eps^R\right]=0,
\end{split}
\en
which follows directly from the Usadel equation (\ref{Eq1}).
The matrix form of $\check{\Delta}$ has a particularly simple form
$\hat{\Delta}=\Delta\,\hat{\Xi}_2$ for the case when $\chi_\eps=0$, which allows one to recast equation (\ref{GSUsadel}) into the following simple form:\cite{Larkin1965,Savich2017}
\beg\label{UsadelAG2}
\left(\eps+\frac{i}{2\tau_{\textrm{s}}}\cosh\vartheta_\eps\right)\sinh\vartheta_\eps=\left(\Delta-\frac{i}{2\tau_{\textrm{s}}}\sinh\vartheta_\eps\right)\cosh\vartheta_\eps.
\en
Introducing functions
\beg\label{Notations}
\tilde{\xi}_{\eps}=\eps+\frac{i}{2\tau_{\textrm{s}}}\cosh\vartheta_\eps, \quad \tilde{\Delta}_\eps=\Delta-\frac{i}{2\tau_{\textrm{s}}}\sinh\vartheta_\eps,
\en
one finds for the complex Nambu angle $\coth\vartheta_\eps={\tilde{\xi}_{\eps}}/{\tilde{\Delta}_\eps}$.
Furthermore, introducing function
\beg\label{etaeps}
\eta_\eps^{R}=\textrm{sign}(\eps)\sqrt{\tilde{\xi}_\eps^2-\tilde{\Delta}_\eps^2}
\en
and using $\cosh\vartheta_\eps={\tilde{\xi}_\eps}/{\eta_\eps^{R}}\equiv g_\eps^{R}$,  
$\sinh\vartheta_\eps={\tilde{\Delta}_\eps}/{\eta_{\eps}^R}\equiv f_\eps^{R}$, one rewrites the Usadel equation (\ref{UsadelAG2}) as follows:
\beg\label{Eq38}
\epsilon=\Delta\coth\vartheta_\eps-\frac{i}{\tau_s}\cosh\vartheta_\eps.
\en
The spectral function (\ref{etaeps})  play an important role in the subsequent analysis.

\begin{figure}
\includegraphics[width=0.850\linewidth]{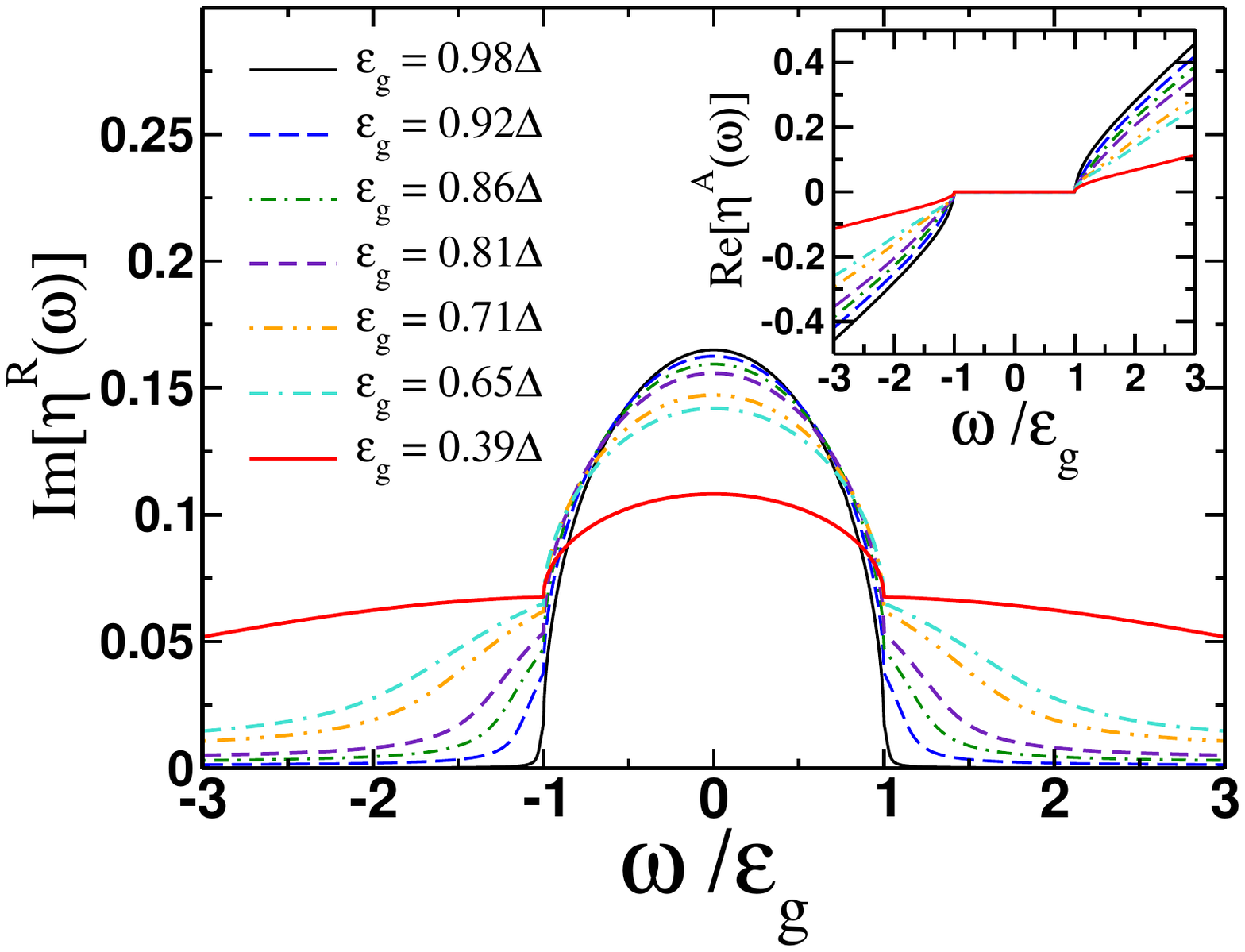}
\caption{Dependence of the real and imaginary part of the function $\eta_\eps^R$, Eq. (\ref{etaRA}), as a function of energy for various values of the pair breaking parameter corresponding to different values of $\veps_g/\Delta$.}
\label{Fig-ETAR}
\end{figure}

Analysis of the Usadel equation (\ref{Eq38}) shows  the quasiparticle gap, $\veps_g$, found from the so-called astroid relation\cite{AG1961,Larkin1965,Savich2017,Kamenev2011}
 \beg
                    \label{astroid}
\veps_g^{2/3} +(1/\tau_s)^{2/3} = \Delta^{2/3}, 
\en
which follows directly from Eq.~(\ref{Eq38}). Here  $\Delta$ is the order parameter, determined self-consistently through Eq.~(\ref{SelfConsistent}) for a given value of the pair-breaking rate $\tau_s^{-1}$.\cite{AG1961}
Moreover, one finds that for $\eps \sim \veps_g$ and $\gamma\ll 1$,  the spectral functions, $\eta_\eps^{R(A)}$, can be approximately written as
\beg\label{etaRA}
\eta_\eps^{R(A)}=\left\{
\begin{aligned}
&\pm \mathrm{sign}(\eps)\sqrt{(\eps\pm i0)^2-\veps_g^2}+i\,\veps_g\gamma^{1/3}, ~ |\eps|\geq\veps_g, \\
&i\sqrt{\veps_g^2-\eps^2}+i\,\veps_g\gamma^{1/3}, \quad |\eps|<\veps_g,
\end{aligned}
\right.
\en
see Fig. \ref{Fig-ETAR}. Here 
\beg\label{gamma}
\gamma=\frac{1}{\tau_s\veps_g}
\en
 is the dimensionless pair-breaking rate.   Note that apart from the extra imaginary part, given by $\veps_g\gamma^{1/3}$, expression (\ref{etaRA}) is analogous to the definition of $\eta_\eps^{R(A)}$ for the case of  a superconductor without pair-breaking, $\tau_s\to\infty$ (see Eq. (\ref{etaRA0}) in Appendix A). Furthermore, Eq. (\ref{etaRA}) is consistent with the results of the exact solution of the fourth order algebraic equation for the function $u_\eps=\tilde{\xi}_\eps/\tilde{\Delta}_\eps$.\cite{Exactu4}

\subsection{Linear analysis}

To derive the expression for the amplitude mode susceptibility we follow the avenue of Refs. [\onlinecite{Moore2017,Kamenev2011, Eremin2024,Dzero2024}]. The derivation calls for allowing a small variation of the order parameter $\hat{\Delta}(\br,t)$ and computing the corresponding corrections to the Keldysh, retarded and advanced components of the $\check{Q}$ matrix by solving the linearized Usadel equation. The resulting expressions are then substituted into the self-consistency condition. This procedure is formally equivalent to the derivation of the expression for the dynamical pairing susceptibility. 

Restricting variations of the order parameter to the longitudinal form:
\beg\label{deltaL}
\delta\hat{\Delta}(\br,t)=\delta\Delta^L(\br,t)\, \hat{\Xi}_2,
\en
provides for the excitation of the collective  Schmid-Higgs amplitude mode.
After performing the Fourier transformation, the corresponding linear correction to the retarded part of the ground state $\check Q$-matrix configuration is given by \cite{Moore2017,Eremin2024,Dzero2024}
\beg\label{deltaQRFin}
\delta\hat{Q}_\eps^R(\bq,\omega)\simeq\frac{\hat{\Lambda}_{\eps+\frac{\omega}{2}}^R\hat{\cal R}_\Delta^R(\eps,\omega)}{\eta_{\eps+\frac{\omega}{2}}^R+\eta_{\eps-\frac{\omega}{2}}^R+iDq^2}\, \delta\Delta_{\bq\omega}^L,
\en
where $D$ is the diffusion coefficient and we adopted the shorthand notation
$\hat{\cal R}_\Delta^R(\eps,\omega)=\left(f_{\eps-\frac{\omega}{2}}^R-f_{\eps+\frac{\omega}{2}}^R\right)\hat{\Xi}_0+\left(g_{\eps+\frac{\omega}{2}}^R+g_{\eps-\frac{\omega}{2}}^R\right)\hat{\Xi}_1$.
Since $\hat{Q}^A=-[\hat{Q}^R]^\dagger$ in full analogy with (\ref{deltaQRFin}), one   has\cite{Moore2017,Eremin2024,Dzero2024}
\beg\label{deltaQAFin}
\delta\hat{Q}_\eps^A(q,\omega)\simeq\frac{\hat{\Lambda}_{\eps+\frac{\omega}{2}}^A\hat{\cal R}_\Delta^A(\eps,\omega)}{\eta_{\eps+\frac{\omega}{2}}^A+\eta_{\eps-\frac{\omega}{2}}^A+iDq^2}\, \delta\Delta_{\bq\omega}^L.
\en
As for the Keldysh component, to separate the linear correction to the quasiparticle spectrum from that to the quasiparticle distribution function, one represents it in the following  form:\cite{Moore2017,Eremin2024,Dzero2024}
\beg\label{Keldysh}
\delta\hat{Q}_\eps^K(q,\omega)=\delta\hat{Q}_\eps^R(q,\omega)t_{\eps-\frac{\omega}{2}}-t_{\eps+\frac{\omega}{2}}\delta\hat{Q}_{\eps}^A(q,\omega)+\delta\hat{g}_\eps^K(q,\omega),
\en
where $t_\eps=\tanh(\eps/2T)$, $T$ is temperature, and
\beg\label{dgKFin}
\delta\hat{g}_\eps^K(q,\omega)\simeq\frac{\hat{\Lambda}_{\eps+\frac{\omega}{2}}^R\left(\hat{\Xi}_2\hat{\Lambda}_{\eps-\frac{\omega}{2}}^A-\hat{\Lambda}_{\eps+\frac{\omega}{2}}^R\hat{\Xi}_2\right)}{\eta_{\eps+\frac{\omega}{2}}^R+\eta_{\eps-\frac{\omega}{2}}^A+iDq^2}\, \delta\Delta_{\bq\omega}^L.
\en
Hereafter we take the limit of very low temperatures, $T\to 0$.

\subsection{Longitudinal Susceptibility}

One inserts now Eqs. (\ref{deltaQRFin})--(\ref{dgKFin}) into the self-consistency equation (\ref{SelfConsistent}).
The resulting linear consistency relation, which will be used for the analysis of the time dependence of the Higgs mode, 
has the  form
$\chi_{\mathrm{SH}}^{-1}(\omega,\bq)=0$, where the inverse longitudinal susceptibility,  $\,\chi_{\mathrm{SH}}^{-1}(\omega,\bq)$,  is given by
\beg\label{MainEq}
\begin{split}
&\chi_{\mathrm{SH}}^{-1}(\omega,\bq)=-\frac{1}{\lambda}+\int\limits_{-\omega_D}^{\omega_D} {d\eps}\left\{\frac{{\cal A}^K({\eps_{+},\eps_{-}})\left(t_{\eps_+}-t_{\eps_-} \right)}{\eta_{\eps_+}^R+\eta_{\eps_-}^A+iDq^2} \right.
\\&\qquad \qquad \left.
+\, \frac{{\cal A}^R({\eps_{+},\eps_{-}})t_{\eps_-}}{\eta_{\eps_+}^R+\eta_{\eps_-}^R+iDq^2} - 
\frac{{\cal A}^A({\eps_{+},\eps_{-}})t_{\eps_+}}{\eta_{\eps_+}^A+\eta_{\eps_-}^A +iDq^2}   
\right\}.
\end{split}
\en
Here $\lambda$ is the dimensionless BCS coupling constant given by
\beg\label{invlam}
\frac{1}{\lambda}=\frac{1}{\Delta}\int\limits_{-\omega_D}^{\omega_D}d\eps\left(f_\eps^R-f_\eps^A\right)t_\eps\, ,
\en
$\omega_D$ is the Debye cutoff frequency, 
$\eps_{\pm}=\eps\pm\omega/2$, and functions ${\cal A}^{K,R,A}({\eps_{+},\eps_{-}})$ are defined as
$${\cal A}^K({\eps_{+},\eps_{-}})=1+g_{\eps_+}^Rg_{\eps_-}^A +f_{\eps_+}^R f_{\eps_-}^A;$$  
$${\cal A}^{R(A)}({\eps_{+},\eps_{-}}) = 1+g_{\eps_+}^{R(A)} g_{\eps_-}^{R(A)} + f_{\eps_+}^{R(A)}f_{\eps_-}^{R(A)}.$$ 
Note that, while the integrals in Eqs.~(\ref{MainEq}), (\ref{invlam}) need to be cut off at a Debye frequency, being taken together they yield a UV convergent integral. Thus expression for the inverse susceptibility, $\chi_{\mathrm{SH}}^{-1}$, is, in fact, cutoff independent. 
 Appendixes A and B  show that for the case $\gamma=0$ one reproduce known expressions for longitudinal susceptibility of superconductors without magnetic impurities in the dirty limit $\tau\Delta\ll1$.\cite{nosov2024}

\begin{figure}
\includegraphics[width=0.8250\linewidth]{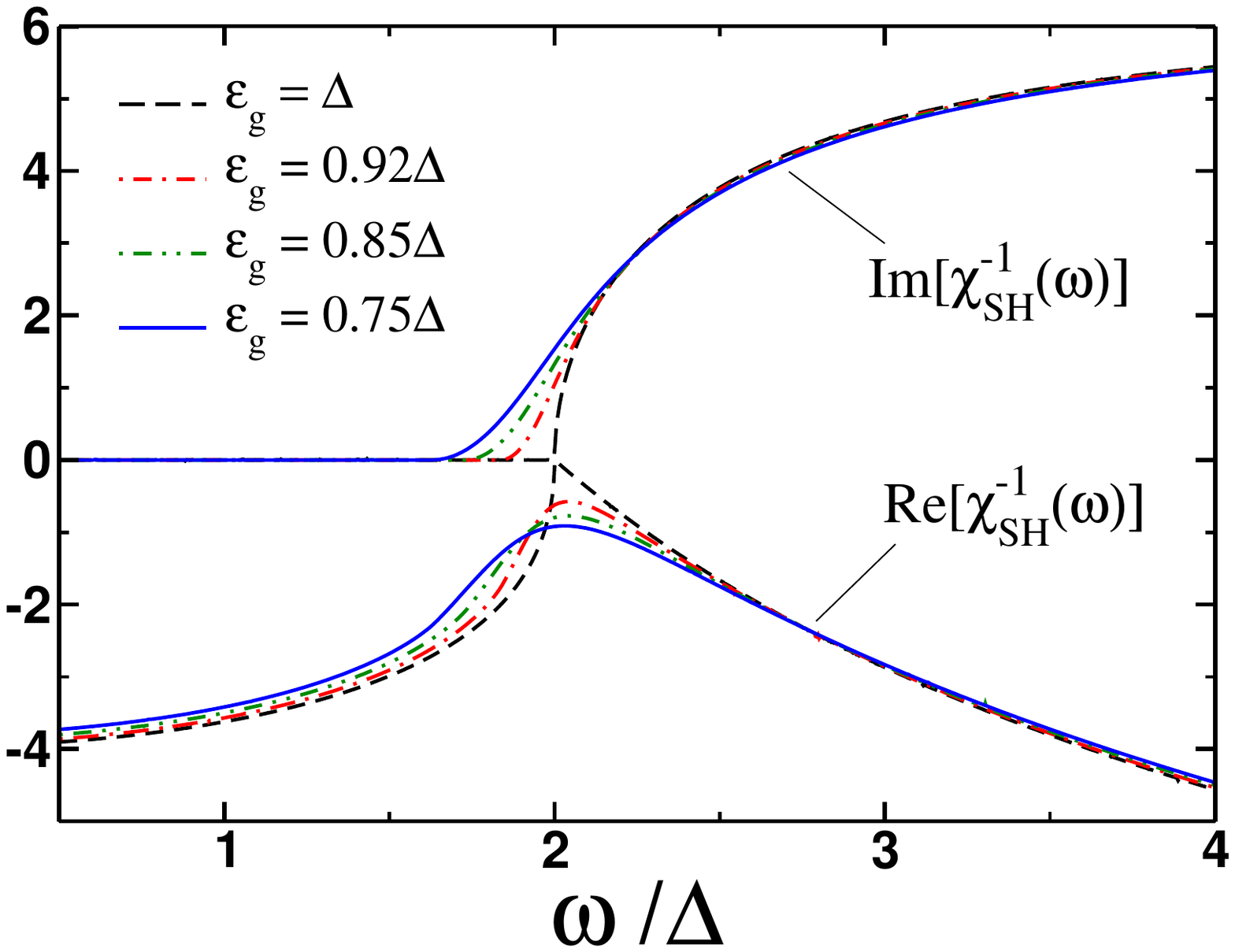}
\caption{Inverse longitudinal susceptibility $\chi^{-1}_{\textrm{SH}}(\omega,\bq=0)$, as a function of frequency for various values of the ratio $\veps_g/\Delta$, in the limit $T\to 0$.}
\label{Fig-OneMore}
\end{figure}

Hereafter we focus  on the behavior of $\chi_{\mathrm{SH}}^{-1}(\omega,\bq)$ at $\bq=0$. Figure~\ref{Fig-OneMore} shows its real and imaginary parts  as  functions of frequency for various values of the pair-breaking parameter, $\gamma$, and thus different gaps, $\veps_g$.  It is noteworthy to emphasize that the real  part of 
$\chi_{\textrm{SH}}^{-1}(\omega)$ touches zero (at $\omega=2\Delta$) only in the presence 
of the time-reversal symmetry, $\gamma=0$. For any finite $\gamma>0$, the real part is always finite. This phenomenon corresponds to the Higgs mode pole being located at the nonphysical sheet of the complex $\omega$ plane, as explained in Refs.~[\onlinecite{ChubukovHiggs2023,nosov2024}].  The imaginary 
part $\mathrm{Im}\,\chi_{\mathrm{SH}}^{-1}(\omega)=0$ for 
$|\omega|<2\veps_g$, reflecting absence of the Landau damping for 
for frequencies below the quasiparticle gap. This statement is only valid at zero temperature, $T=0$, while at a finite temperature a small tail  for $|\omega|<2\veps_g$ appears due to the inter-band processes.

 Given expressions for the functions $\eta_\eps^{R(A)}$, Eq.~(\ref{etaRA}),  one notices that expressions for the susceptibility are the same as for time-reversal symmetric superconductors with the substitutions: $\Delta\to\eps_g$ and 
 $Dq^2\to Dq^2+2\veps_g\gamma^{1/3}$. 
 One can thus benefit from approximate expressions for  $\chi_{\textrm{SH}}(\omega,\bq)$ derived for the time-reversal unbroken  superconductors\cite{nosov2024} (see also Appendices A and B).  Here we limit ourselves to the region of small wavenumbers $q\lesssim \xi^{-1} \gamma^{1/6}$, where $\xi=\sqrt{D/\veps_g}$ is the superconducting coherence length.  As a result, for $\omega \sim 2\eps_g$ the real part of $\chi_{\textrm{SH}}^{-1}(\omega)$ is given by
\beg\label{RechiSHgamma}
\begin{split}
\mathrm{Re}[\chi_{\textrm{SH}}^{-1}(\omega)]\approx&\,\,\gamma^{1/3}\left\{\log\left(\frac{e^2}{\gamma^{2/3}}\right)
-\log \sqrt{\zeta_\omega}\right.\\&\left.-\sqrt{1+4\zeta_\omega}\log\left(\frac{1+\sqrt{1+4\zeta_\omega}}{2\sqrt{\zeta_\omega}}\right)\right\},
\end{split}
\en
where
\beg\label{zetaw}
\zeta_\omega=\frac{|\omega-2\veps_g|}{\veps_g\gamma^{2/3}}. 
\en
Similarly, for the imaginary part of $\chi_{\textrm{SH}}^{-1}(\omega)$ one obtains:
\beg\label{ImchiSHgamma}
\begin{split}
\textrm{Im}\left[\chi_{\mathrm{SH}}^{-1}(\omega)\right]&\approx  
\pi\gamma^{1/3}\,\vartheta(\omega-2\veps_g)\left\{\sqrt{1+4\zeta_\omega}-1\right\}.
\end{split}
\en
It is straightforward to verify that in the limit $\gamma\to0$ we recover the corresponding expressions for $\mathrm{Re}[\chi_{\textrm{SH}}^{-1}(\omega)]$ and $\mathrm{Im}[\chi_{\textrm{SH}}^{-1}(\omega)]$ in the time-reversal unbroken  superconductor (see Appendix A and Appendix B for details). The non-analytic behavior of Eqs.~(\ref{RechiSHgamma}) and (\ref{ImchiSHgamma}) at $\omega=2\veps_g$ is of central importance.  

\begin{figure}
\includegraphics[width=0.8250\linewidth]{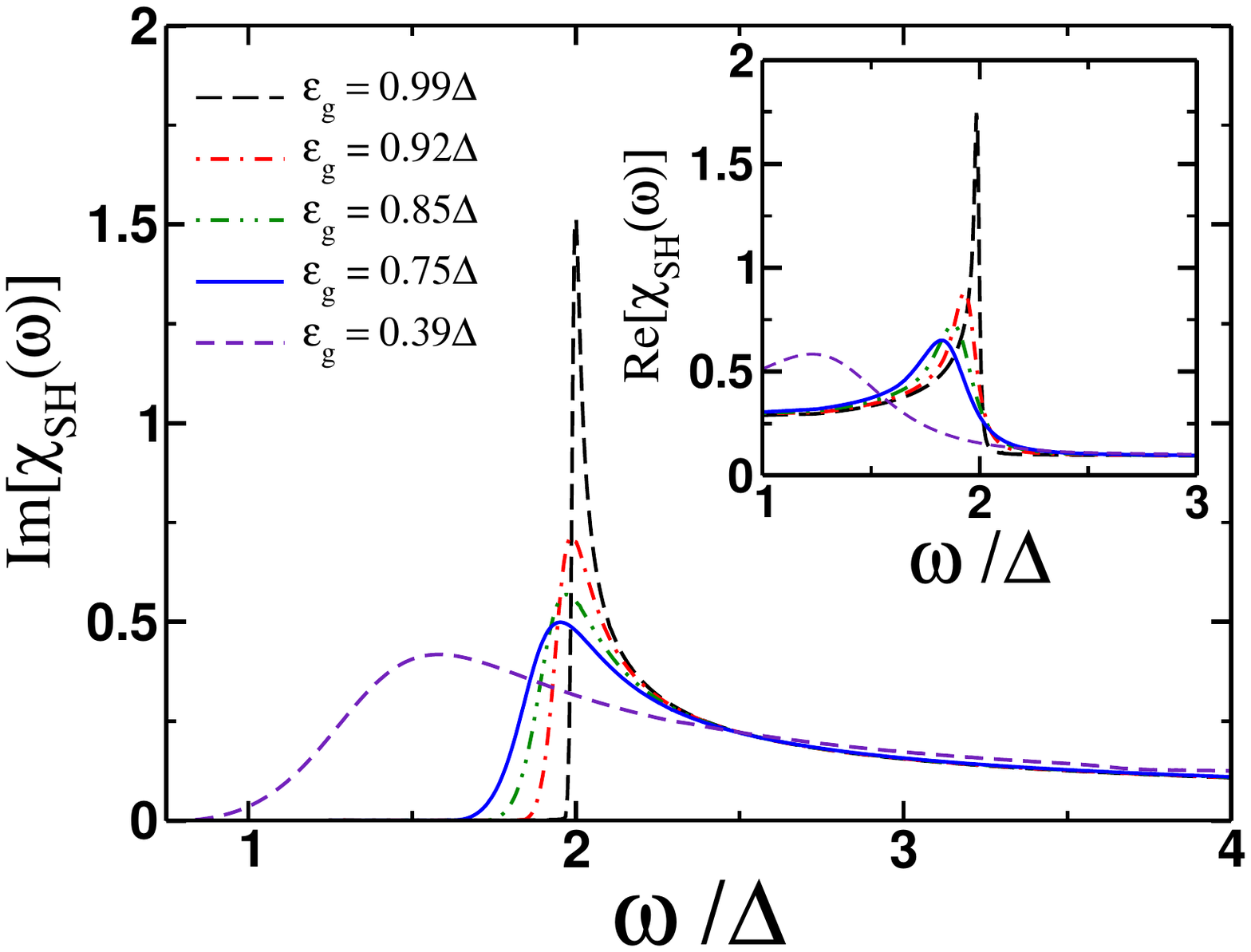}
\caption{Dependence of the real and imaginary parts of the amplitude mode susceptibility $\chi_{\textrm{SH}}(\omega,\bq=0)$, as a function of energy for various values of the ratio $\veps_g/\Delta$ in the limit $T\to0$.}
\label{Fig-ImRechiSH}
\end{figure}

Figure \ref{Fig-ImRechiSH} shows real and imaginary parts of $\chi_{\mathrm{SH}}(\omega,\bq=0)$ for different values of the pair-breaking parameter, $\gamma$.
Note that both functions $\textrm{Re}[\chi_{\textrm{SH}}(\omega)]$ and $\textrm{Im}[\chi_{\textrm{SH}}(\omega)]$ exhibit a peak as a function of frequency at $\omega=2\omega_{\textrm{res}}$. Employing Eqs.~(\ref{RechiSHgamma}) and (\ref{ImchiSHgamma}), one estimates it for $\gamma\ll 1$ as
\beg\label{wres}
\omega_{\textrm{res}}\approx \veps_g\left[1+\frac{\gamma^{2/3}}{4\pi^2}\log^2\left(\frac{e^2}{\gamma^{2/3}}\right)\right],
\en
putting the peak frequency within the window  $\veps_g< \omega_{\textrm{res}} \lesssim \Delta$. This peak signifies absorption maximum for an energy source, which is coupled to the longitudinal fluctuations of the order parameter. In linear order electromagnetic radiation couples only to transverse fluctuations, which exhibit plasmon gap in superconductors with charged Cooper pairs. However,  two-phonon process (associated with the square of the vector potential in the Usadel equation (\ref{Eq1})) are coupled with the longitudinal mode.\cite{Moore2017,Eremin2023,Dzero2024,Li2025} As a result, one expects to see a maximum in two-photon absorption of an IR radiation with $\omega = \omega_{\textrm{res}}$. Notice that, as long as $\omega_{\textrm{res}} < 2\veps_g$, the one-photon absorption by quasiparticle excitations is still absent.     


\section{Dynamics of the amplitude Schmid-Higgs mode}

We now turn to a discussion of the time-dependence of the space homogeneous, $\bq=0$,  amplitude mode. The time dependence of $\chi_{\mathrm{SH}}(t)$ describes the system's response to a small sudden perturbation. Indeed, as it can be checked by a direct calculation, in both clean and disordered conventional superconductors with the order parameter $\Delta_0$, the susceptibility is $\chi_{\mathrm{SH}}(t)\sim\sin(2\Delta_0 t)/\sqrt{\Delta_0t}\,$, at times $t\Delta_0\gg 1$. This is  in agreement with the long-time dynamics of the pairing amplitude $\Delta(t)=\Delta_{\textrm{s}}[1+a\cos(2\Delta_0t+\pi/4)/\sqrt{\Delta_0 t}\,]$ following a sudden perturbation such as  small change of the pairing strength.\cite{VolkovKogan1973} In passing we note that $\Delta_{\textrm{s}}$ in the expression above is given by $\Delta_{\textrm{s}}=\Delta_0+{\delta\Delta^2}/{\Delta_0}$ where $\delta\Delta=\Delta_0'-\Delta_0$, $\Delta_0'$ denotes the value of the order parameter in the new equilibrium state, which means that within the linear analysis amplitude mode asymptotes its value in equilibrium.\cite{Yuzbashyan2013}
For strong enough quenches the order parameter periodically oscillates with time,\cite{Spivak2004,Levitov2006,Yuzbashyan2006,Yuzbashyan2008,Levitov2007,Yuzbashyan2013} however to capture such a dynamics within the present formalism one needs to go beyond the linear analysis. 

\begin{figure}
\includegraphics[width=0.9750\linewidth]{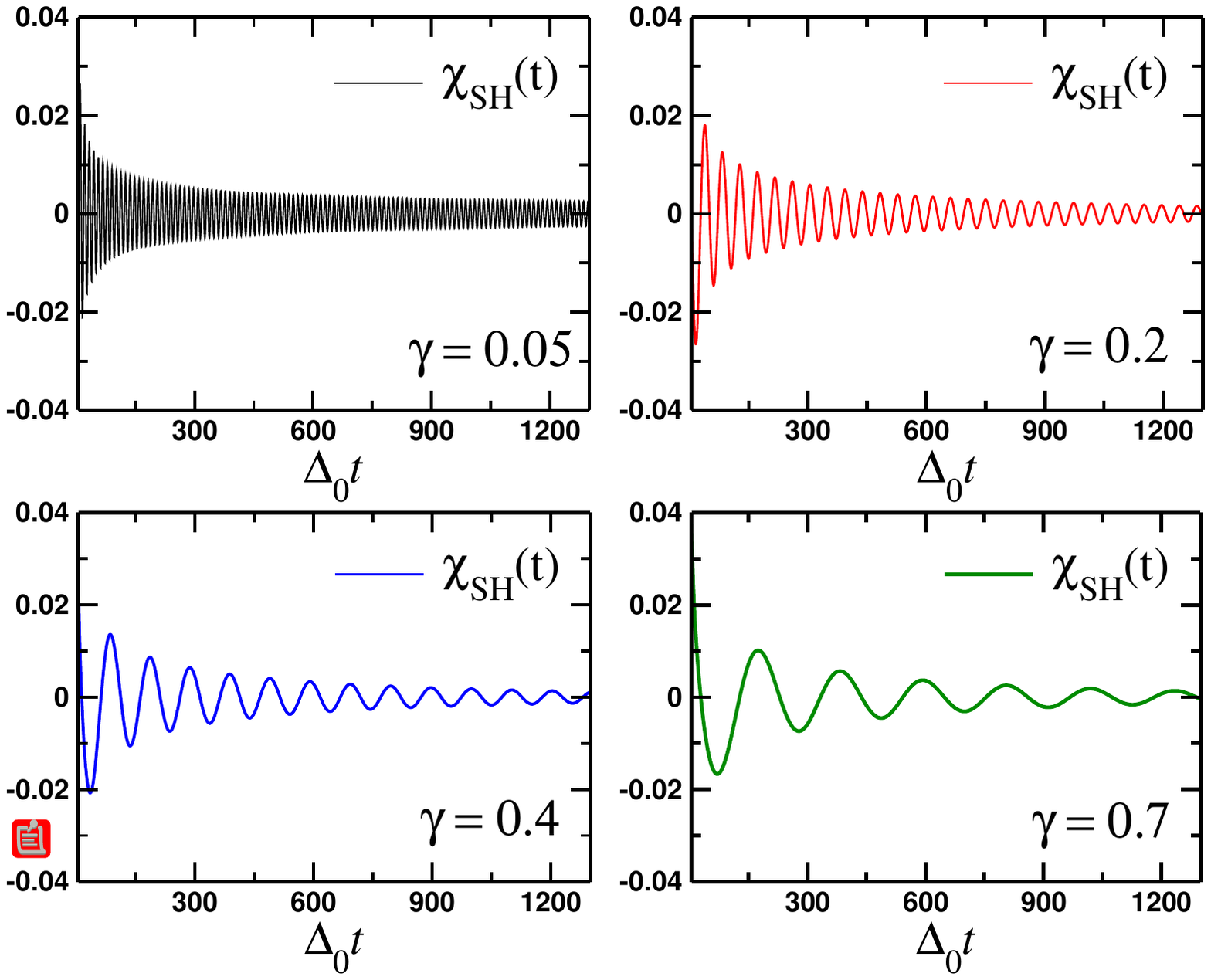}
\caption{Longitudinal mode response on a sudden quench perturbation at $t=0$, evaluated for different values of the dimensionless parameter $\gamma=1/\tau_s\veps_g$. Here $\Delta_0$ is the superconducting pairing gap for $\tau_s\to\infty$. At long times, such that $t\gg\tau_s \gamma^{1/3}$, the susceptibility asymptotes to zero as $\sim 1/t^2$.}
\label{Fig-chiSHt}
\end{figure}

The time dependence of the Schmid-Higgs susceptibility at $\bq=0$ can be found by evaluating the Fourier transform of $\chi_{\textrm{SH}}^R(\omega)$.
Given its analytical properties, one can write it as:\cite{nosov2024}
\beg\label{chiSHtMain}
\chi_{\textrm{SH}}(t)= -\mathrm{Re}\int\limits_{2\veps_g}^\infty \frac{d\omega}{\pi}\,\, \chi_{\mathrm{SH}}^R(\omega)\left(e^{i\omega t}-e^{-i\omega t}\right).
\en
It is now straightforward to numerically evaluate this integral using the approximate expressions (\ref{RechiSHgamma}) and (\ref{ImchiSHgamma}).
In Fig. \ref{Fig-chiSHt} we present the results of this calculation for different values of the dimensionless parameter $\gamma = 1/(\tau_s\veps_g)$. To further analyze Eq.~(\ref{chiSHtMain}),  one shifts the integration variable to the imaginary axis $\omega=2\veps_g+iy$ in the first term on its right hand side, while in the second term one makes the similar change $\omega=2\veps_g-iy$. It then follows:\cite{nosov2024}
\beg\label{chiSHrotated}
\begin{split}
\chi_{\textrm{SH}}(t)&=\mathrm{Im}\left\{ e^{2i\veps_gt}\int\limits_{0}^\infty \frac{dy}{\pi}\,\, \chi_{\mathrm{SH}}^R(2\veps_g+iy)\, e^{-yt}  \right. \\&\qquad + \left. e^{-2i\veps_gt}\int\limits_{0}^\infty \frac{dy}{\pi}\,\, \chi_{\mathrm{SH}}^R(2\veps_g-iy)\, e^{-yt}\right\}.
\end{split}
\en
It is easy to check then using expressions from Appendix B,  that in the limit $\tau_s\to\infty$, function $\chi_{\textrm{SH}}(t)$ behaves as $\sin(2\Delta t)/\sqrt{t}$. Furthermore, for finite $\tau_s$ at long times the integral accumulates in the region of small $y\lesssim 1/t$, which allows one  expand $\chi_{\mathrm{SH}}(2\veps_g\pm iy)$ in powers of $y$. Since $\mathrm{Im}\, \chi_{\mathrm{SH}}(2\veps_g)=0$, cf. Eq.~(\ref{ImchiSHgamma}), the first non-vanishing contribution appears in the linear  order in  $y$. As a result,  for finite values of $\gamma$ and long times one finds 
\beg\label{chiSHrotatedFinal}
\chi_{\textrm{SH}}(t) =\frac{\vartheta(t)}{\gamma \log^2(e/\gamma^{1/3})} \,  \frac{\sin(2\veps_gt)}{\veps_g t^2}\, , 
\en
where we used Eqs.~(\ref{RechiSHgamma}) and (\ref{ImchiSHgamma}) for $\chi_{\textrm{SH}}(\omega)$. 

To understand the characteristic crossover time between the two regimes, one notices that the aforementioned expansion of $\chi_{\mathrm{SH}}(2\veps_g\pm iy)$ works as long as 
$y\lesssim \Delta-\veps_g \approx \veps_g\gamma^{2/3}$, i.e. $\zeta_\omega \lesssim 1$ cf. Fig.~\ref{Fig-OneMore} and Eq.~(\ref{zetaw}). Therefore the validity of Eq.~(\ref{chiSHrotated}) is justified for $t\gg 1/\veps_g\gamma^{2/3}$.
Notice that for small $\gamma\ll 1$, the crossover time is 
very different from naive estimates: $\Delta^{-1}\ll 1/\veps_g\gamma^{2/3} \ll \tau_s$. In the opposite limit of short times, 
$t\ll 1/\veps_g\gamma^{2/3}$, the response function is given by
\beg\label{chiSHrotatedFinal-short}
\chi_{\textrm{SH}}(t) \propto  \vartheta(t)\, \frac{\sin(2\veps_gt)}{\sqrt{t/\veps_g}}.
\en
Equations (\ref{chiSHrotatedFinal}) and (\ref{chiSHrotatedFinal-short}) are the main result of this paper. They are in parametric agreement with each other at time $t\approx 1/\veps_g\gamma^{2/3}$ (up to a logarithmic factor) and are consistent with the numerical calculations, Fig. \ref{Fig-chiSHt}.
This behavior is reminiscent to that found recently for the long time dynamics of the finite momentum Schmidt-Higgs susceptibility, $\chi_{\textrm{SH}}(t,\bq)$, in conventional superconductors under  sudden spatially periodic perturbations.\cite{nosov2024}

\section{Discussion and Conclusions}

\begin{figure}
\includegraphics[width=0.850\linewidth]{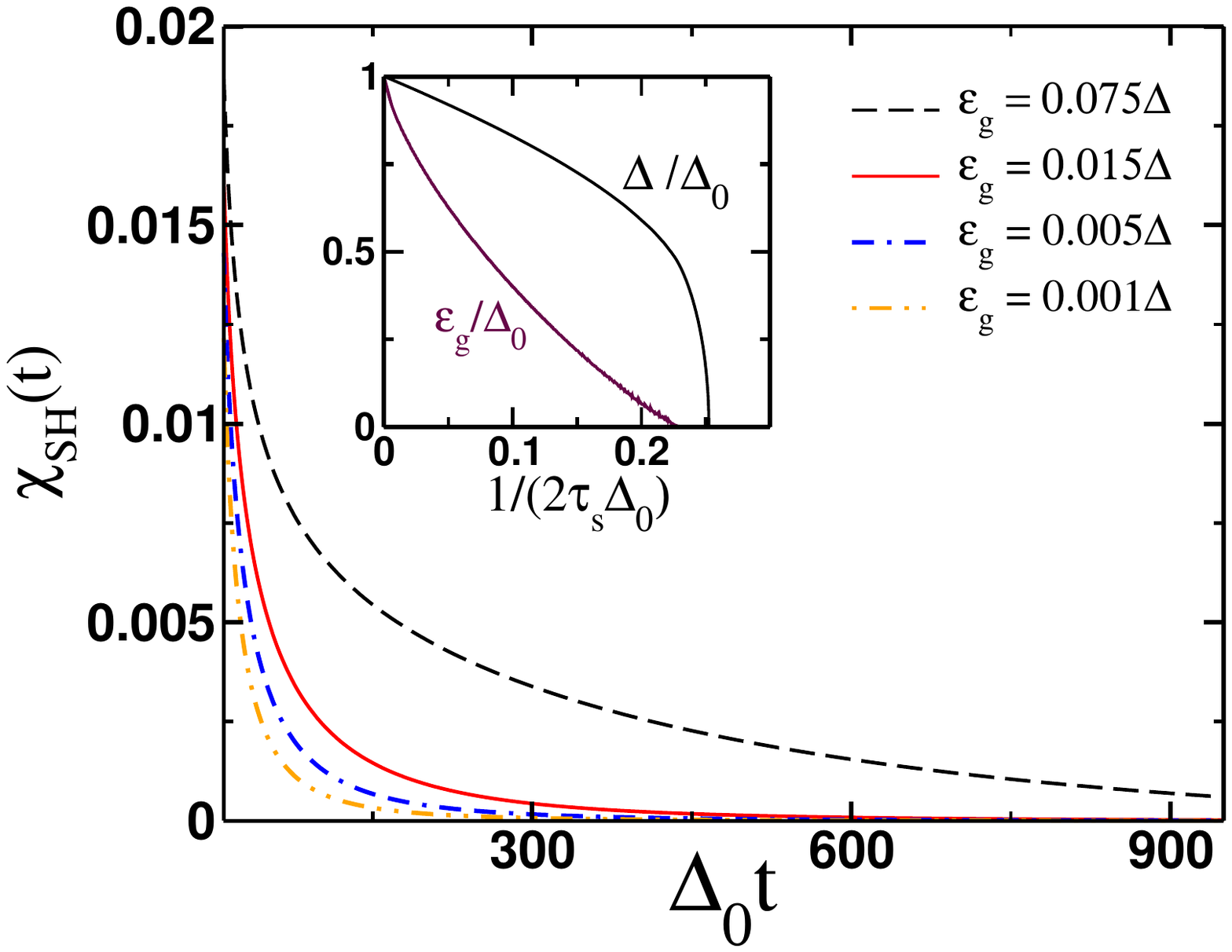}
\caption{Time evolution of the Schmid-Higgs susceptibility evaluated for $\tau_s\Delta\approx1$ when a superconductor is close to the gapless state $\veps_g/\Delta=0$ (inset). Here $\Delta_0$ is the superconducting pairing gap for $\tau_s\to\infty$. At long times, such that $t\gg\tau_s$, the susceptibility asymptotes to zero as $1/t^2$. Inset: dependence of the quasiparticle (half) gap, $\veps_g$, and the order parameter, $\Delta$, on the pair-breaking rate, $1/\tau_s$. The gapless superconductivity regime is seen for $0.228 < (2\tau_s\Delta_0)^{-1}< 0.25$. }
\label{Fig-gapless}
\end{figure}

One aspect of the problem that we did not discuss so far concerns dynamics of the amplitude mode in the gapless state, $\veps_g = 0$, while $\Delta \neq 0$. As  follows from the expression (\ref{astroid}) for $\veps_g$, the gapless state appears when the pair-breaking rate exceeds the self-consistently found order parameter,  $1/\tau_s\geq \Delta$, see inset in Fig.~\ref{Fig-gapless}. It is worth mentioning that the transition to the gapless state is accompanied by the topological Lifshitz transition.\cite{Yerin2022a,Yerin2022b} At such a transition the non-analytic points at real $\omega =\pm 2\veps_g$ collide at $\omega=0$ and move into the complex frequency plane in accordance with solutions of Eq.~(\ref{astroid})  (only singularities in the lower (upper) half plane appear in $\chi_{\mathrm SH}^{R(A)}(\omega)$).  As a result, the longitudinal susceptibility is analytic along the entire real axis of $\omega$. This makes its Fourier transform, $\chi_{\textrm{SH}}(t)$, an exponentially decaying function of time. The corresponding exponent is given by the imaginary part of $2\veps_g$, determined  from Eq.~(\ref{astroid}). The absence of a long time algebraic decay  is consistent with $1/\gamma$ factor in Eq.~(\ref{chiSHrotatedFinal}), as $\gamma\to\infty$ at approaching the gapless regime.  It is also worth mentioning that a finite temperature, $T>0$, also removes non-analytic points at  $\omega =\pm 2\veps_g$ and thus results in the exponential decay of Schmid-Higgs oscillations at $t>1/(\pi T)$. 

In conclusion, we have studied dynamics of the collective Schmid-Higgs amplitude mode in disordered superconductors with a finite pair-breaking rate, $1/\tau_s$. The latter is achieved by either an in-plane magnetic field for thin film superconductors, or  a small concentration of pair-breaking centers such as weak magnetic impurities. 
Without pair-breaking, the frequency of the longitudinal Schmid-Higgs mode at ${\bf q} =0$, is given by $2\Delta_0$, which is exactly the quasiparticle gap and simultaneously (twice of) the superconducting order parameter. Being situated right at the edge of the continuous quasiparticle spectrum, leads to a slow algebraic decay of the longitudinal oscillations of the order parameter with time, $\sim t^{-1/2}$.  In presence of a pair-breaking mechanism, the mean-field quasiparticle (half) gap, $\veps_g$, is different from the superconducting order parameter, $\Delta$ and is, quite generally, smaller, $\veps_g<\Delta$. This could lead to a situation when the longitudinal mode falls inside the quasiparticle continuum with an expected exponential damping. 

We showed that the situation is more nuanced. The maximum of the imaginary part of the linear susceptibility, given by Eq.~(\ref{wres}), indeed falls inside the quasiparticle continuum. This leads to a broad peak in the two-photon absorption spectrum, with its maximum being at $\omega_{\mathrm{res}}>\veps_g$. However, response to a sudden perturbation, or quench is very different. It exhibits algebraically decaying oscillations with $\omega=2\varepsilon_g$. At long times such an algebraic decay is characterized by $t^{-2}$ law (cf. with $t^{-1/2}$ decay of the $\bq=0$ mode, in superconductors without pair-breaking, where the corresponding peak in the absorption spectrum is highly asymmetric with the sharp threshold at $2\Delta_0$ and $(\omega-2\Delta_0)^{-1/2}$ singularity above it). Notice also that the algebraic decay pattern  found here for $\bq =0$ and a finite pair-breaking rate, is similar to decay of finite $\bf q$ sudden perturbations in conventional superconductors, found  
recently  in Ref.~[\onlinecite{nosov2024}]. 
The correspondence rule appears to be $\gamma \leftrightarrow  (\xi q)^6$, where $\xi$ is the superconducting coherence length and $\gamma=1/(\tau_s\veps_g)$ is a dimensionless pair-breaking rate.

\section{Acknowledgements}
We are grateful to  Igor Burmistrov, Andrey Chubukov, Pavel Nosov, and Yahor Savich  for illuminating discussions. A.K. was financially supported by the NSF grant DMR-2338819.

\begin{appendix}
\section{Imaginary part of  $\chi_{\mathrm{SH}}^{-1}(\omega,\bq)$}
In this Section we provide the results of the calculation of the imaginary part of the inverse amplitude mode susceptibility $\chi_{\mathrm{SH}}^{-1}(\omega,\bq)$ for the case when there are no magnetic impurities for in this case, as it will become clear in what follows,  $\chi_{\mathrm{SH}}^{-1}(\omega,\bq)$ can be computed analytically to a very good accuracy.

From the definition of the function $\chi_{\mathrm{SH}}^{-1}(\omega,\bq)$ (see the discussion below Eq. (\ref{MainEq}) in the main text), it is clear that the imaginary part of this function will be given by the sum of the imaginary parts of the three integrals whose kernels are defined by the functions ${\cal A}^K(\eps_{+},\eps_{-})$ and ${\cal A}^{R(A)}(\eps_{+},\eps_{-})$ correspondingly. 

All the integrals can be evaluated by taking into account that we will be primarily interested in finding the pairing susceptibility for frequencies $\omega=2\Delta+\nu$ with $\nu\ll \Delta$.
Let us first consider the case when $\bq=0$.
Given the definitions
\beg\label{etaRA0}
\eta_\eps^{R(A)}=\left\{
\begin{aligned}
&\pm\mathrm{sign}(\eps)\sqrt{(\eps\pm i0)^2-\Delta^2}, \quad |\eps|\geq \Delta, \\
&i\sqrt{\Delta^2-\eps^2}, \quad |\eps|<\Delta
\end{aligned}
\right.
\en
it will be convenient to adopt the limit of low temperatures $T\to 0$. We also take into account that the functions under the integrals are even functions of energy and the integration can be performed over the positive values of energy. In the first integral involving ${\cal A}^K(\eps_{+},\eps_{-})$ the integration is performed over the segment $\eps\in[0,\Delta+\nu/2]$ and we use the following approximation 
\beg\label{ApproxetaR}
\sqrt{\eps+\frac{\nu}{2}}\sqrt{\eps+\frac{\nu}{2}+2\Delta}\approx\sqrt{2\Delta}\sqrt{\eps+\frac{\nu}{2}}.
\en
The remaining two integrals can be approximated in the same way. 
After a straightforward calculation we obtain
\beg\label{Result4q0}
\mathrm{Im}[\chi_{\mathrm{SH}}^{-1}(\omega,0)]\approx-\left(\frac{2\pi}{\omega}\right)\vartheta(|\omega|-2\Delta)\sqrt{\omega^2-4\Delta^2}.
\en
Next, we keep $\bq$ finite and repeat the calculation again by taking into account that $\nu=|\omega-2\Delta|\ll \Delta$, $Dq^2\ll \Delta$ while keeping the ratio $\nu/Dq^2$ of the order $O(1)$. Adopting the same approximation (\ref{ApproxetaR}) as before we find
\beg\label{invchiSHFin}
\begin{split}
\textrm{Im}\left[\chi_{\mathrm{SH}}^{-1}(\omega,\bq)\right]&\approx  
\pi\textrm{sign}(\omega)\vartheta(|\omega|-2\Delta)\\&\times\left(\frac{Dq^2}{\Delta}\right)\left\{1-\sqrt{1+4|z_\omega|}\right\}.
\end{split}
\en
Here we have introduced new dimensionless variables
$z_\omega=(\frac{\omega}{\Delta}-2)/{\xi^4 q^4}$ and $\xi^2q^2=Dq^2/\Delta$ ($\xi$ is the length scale of the order of the coherence length). Expressions (\ref{Result4q0}) and (\ref{invchiSHFin}) will be used in the calculation of the function 
$\mathrm{Re}[\chi_{\textrm{SH}}(\omega,\bq)]^{-1}$ below.
\section{Real part of  $\chi_{\mathrm{SH}}^{-1}(\omega,\bq)$}
The real part of the function $\chi_{\mathrm{SH}}^{-1}(\omega,\bq)$ can be computed by evaluating the principal value of the following integral
\beg\label{PIReChi}
\begin{split}
\mathrm{Re}[\chi_{\textrm{SH}}(\omega,\bq)]^{-1}=\dashint\limits_{-\infty}^\infty\frac{d\veps}{\pi}&\left\{\frac{\mathrm{Im}[\chi_{\textrm{SH}}(\veps,\bq)]^{-1}}{\veps-\omega}\right.\\&\left.+\frac{\pi}{2}\frac{\vartheta(|\veps|-2\Delta)}{\sqrt{\veps^2-4\Delta^2}}\right\}.
\end{split}
\en
The second term under the integral guarantees that function $\mathrm{Re}[\chi_{\textrm{SH}}(\omega=2\Delta,\bq)]^{-1}$ vanishes at $\bq=0$. In this Section - just like in the previous one - we consider the case $\tau_s\to\infty$. 

The calculation proceeds in several steps. The first step consists in computing $\mathrm{Re}[\chi_{\textrm{SH}}(\omega,\bq=0)]$. With the help of (\ref{Result4q0}) and after simple algebra we have
\beg\label{ReChiSHq0}
\begin{split}
\mathrm{Re}[\chi_{\textrm{SH}}(\omega,\bq=0)]^{-1}=(\delta^2-1)\dashint\limits_{1}^\infty\frac{d\veps}{(\veps^2-\delta^2)\sqrt{\veps^2-1}}.
\end{split}
\en
where $\delta=\omega/2\Delta$. This integral needs to be evaluated separately for situations when $\delta<1$ and $\delta>1$. As a result of the energy integration we obtain:
\beg\label{ReChiepsq0}
\begin{aligned}
&\mathrm{Re}[\chi_{\textrm{SH}}(\omega,0)]^{-1}=\vartheta(1-\delta)\frac{\sqrt{1-\delta^2}}{\delta}\arcsin\delta\\&+\vartheta(\delta-1)\frac{\sqrt{\delta^2-1}}{\delta}\\&\times
\log\left[1+4\delta\sqrt{\delta^2-1}(1-2\delta^2)+8\delta^2(\delta^2-1)\right]^{1/4}.
\end{aligned}
\en
Using the fact that the real part of function $\chi_{\textrm{SH}}^{-1}(2\Delta,0)$ is identically zero, we subtract from both sides of  
(\ref{ReChiSHq0}) the integral representation for $\mathrm{Re}[\chi_{\textrm{SH}}(2\Delta,0)]^{-1}$. It then follows
\beg\label{ReChiSH2DqA}
\begin{split}
&\mathrm{Re}[\chi_{\textrm{SH}}(2\Delta,\bq)]^{-1}\\&=\frac{1}{\pi}\dashint\limits_{2\Delta}^\infty\frac{2\veps d\veps}{\veps^2-4\Delta^2}
\left\{\mathrm{Im}[\chi_{\textrm{SH}}(\veps,\bq)]^{-1}-\mathrm{Im}[\chi_{\textrm{SH}}(\veps,0)]^{-1}\right\}.
\end{split}
\en
In order to evaluate this integral we introduce a new integration variable $y=\veps/\Delta-2$ and perform the integration by parts. By taking into account that the integral accumulates in the region where $y\ll 1$ we approximately find
\beg\label{ReChiSH2DqFin}
\begin{split}
\mathrm{Re}[\chi_{\textrm{SH}}(2\Delta,\bq)]^{-1}&\approx\dashint\limits_0^\infty\left(\frac{\log(y)}{\sqrt{y+(\xi^2q^2/2)^2}}-\frac{\log(y)}{\sqrt{y}}\right)dy\\&=2\xi^2q^2\log\left(\frac{e}{\xi^2 q^2}\right).
\end{split}
\en

We can now proceed with the calculation of $\mathrm{Re}[\chi_{\textrm{SH}}(\omega,\bq)]^{-1}$. We have
\beg\label{ComputeReChiSH}
\begin{split}
&\mathrm{Re}[\chi_{\textrm{SH}}(\omega,\bq)]^{-1}=\mathrm{Re}[\chi_{\textrm{SH}}(2\Delta,\bq)]^{-1}\\&+\frac{1}{\pi}\dashint_{2\Delta}^\infty \mathrm{Im}[\chi_{\textrm{SH}}(\veps,\bq)]^{-1}\left(\frac{2\veps}{\veps^2-\omega^2}-\frac{2\veps}{\veps^2-4\Delta^2}\right)d\veps. 
\end{split}
\en
This integral can be computed by invoking the integration by parts. Taking into account that the main contribution to the resulting integral accumulates in the region $\veps\sim 2\Delta$, it follows
\begin{widetext}
\beg\label{ComputeReChiSH1}
\begin{split}
&\frac{1}{\pi}\dashint_{2\Delta}^\infty \mathrm{Im}[\chi_{\textrm{SH}}(\veps,\bq)]^{-1}\left(\frac{2\veps}{\veps^2-\omega^2}-\frac{2\veps}{\veps^2-4\Delta^2}\right)d\veps
\approx\frac{1}{2}\dashint_0^\infty\frac{dz}{\sqrt{z+\overline{q}^4}}\log\left(\frac{|y+2-\frac{\omega}{\Delta}|}{y}\right).
\end{split}
\en
Here $\overline{q}^4=(Dq^2/2\Delta)^2$ and on the last step we again used the approximation similar to (\ref{ApproxetaR}).
After simple algebraic manipulations, the integration results in the following expression for the function $\mathrm{Re}[\chi_{\textrm{SH}}(\omega,\bq)]^{-1}$:
\beg\label{ReChiSHinvFinFin}
\begin{split}
&\mathrm{Re}[\chi_{\textrm{SH}}(\omega,\bq)]^{-1}\approx2\left(\frac{Dq^2}{\Delta}\right)\log\left[\frac{e\Delta}{Dq^2}\right]
-\left(\frac{Dq^2}{\Delta}\right)\left\{\frac{1}{2}\log z_\omega+\sqrt{1+4z_\omega}\log\left(\frac{1+\sqrt{1+4z_\omega}}{2\sqrt{z_\omega}}\right)\right\}.
\end{split}
\en
\end{widetext}
Expressions (\ref{invchiSHFin},\ref{ReChiSHinvFinFin}) can now be used to estimate the real part of the amplitude mode susceptibility for the case when weak magnetic impurities are present in a superconductor. Finally, it is worth noting here that  expressions for $\mathrm{Re}[\chi_{\textrm{SH}}(\omega,\bq)]^{-1}$ and $\mathrm{Im}[\chi_{\textrm{SH}}(\omega,\bq)]^{-1}$ listed above match (up to an overall factor of $4$) the corresponding expressions given in Ref. \onlinecite{nosov2024}.

\end{appendix}


\begin{thebibliography}{10}
\bibitem{Ref41}
 By using the bare correlation functions instead of the ones 'dressed'
  by disorder in the collision integral one will unavoidably miss the emergence
  of the energy scale $\veps_g$ corresponding to the single particle excitation
  threshold. This circumstance is not surprising given that $\veps_g$ is a
  non-analytic function of the dimensionless pair breaking scattering rate
  $\zeta=1/(\tau_s\Delta)$. In fact,
  $d\veps_g/d\zeta=-\Delta\sqrt{1-\zeta^{2/3}}/\zeta^{1/3}$ and exactly because of
  this non-analytic behavior it is not amenable to any non-self-consistent
  perturbative approach.
\end{thebibliography}

\end{document}